# Entanglement and separability of qubits systems related to measurement processes, Hilbert-Schmidt (HS) decompositions and general Bell stats

Y. Ben-Aryeh

*Physics Department Technion-Israel Institute of Technology, Haifa, 32000, Israel*

E-mail: phr65yb@physics.technion.ac.il

ABSTRACT

Separability and entanglement for n-qubits systems are quantified by using Hilbert-Schmidt (HS) decompositions, in which the density matrices are decomposed into various terms representing certain 1-qubit, 2-qubits…, n-qubits measurements. The present method is more general than previous methods for bipartite systems, as it can be used for quantification of entanglement for large n-qubits systems ( $n \geq 3$ ). We demonstrate the use of the present method by analyzing 3-qubits GHZ states and 3-qubits general Bell-states produced by a certain multiplications of Braid operators, operating on the computational basis of states. Quantum correlations are obtained by measuring all qubits of these systems, while a measurement of a part of these systems gives only classical correlations. Quantification of entanglement, for these systems, is given by the use of HS parameters.

PACS: 03.67.Lx; 42.50.Ex.

## 1. Introduction

Entanglement of qubits systems is at the core of quantum computation field [1-3]. Therefore, it is of utmost importance to quantify entanglement in such systems and to have a definite criterion when such systems become separable. While there are advanced methods for treating entanglement in bipartite systems [4-10], the procedure for quantifying entanglement and separability for larger n-qubits systems ( $n > 2$ ) becomes more complex and less definite. I show in the present work a special method for quantifying entanglement for large n-qubits systems, based on Hilbert-Schmidt (HS) decompositions [11].

I use in the present work the following definition of separability for bipartite system: A density state $\rho$ on Hilbert space $H_A \otimes H_B$ where $A$ and $B$ are the two parts of a bipartite system is defined as *non-entangled-separable* if there exist density operators $\rho_A{}^{(j)}, \rho_B{}^{(j)}$ and $p_j \geq 0$ with $\sum_j p_j = 1$ such that

$$\rho = \sum_j p_j \rho_A^{(j)} \otimes \rho_B^{(j)} \quad . \tag{1}$$

The interpretation for such definition is that for bipartite separable states the systems described by $\rho_A^{(j)} \otimes \rho_B^{(j)}$ are completely independent of each other. In the present work I use the following generalized definition of separability for a system composed of three parts, $A, B$ and $C$ :

$$\rho = \sum_j p_j \rho_A^{(j)} \otimes \rho_B^{(j)} \otimes \rho_C^{(j)} \quad . \tag{2}$$

Consequent definitions can be given for larger n-qubits systems ($n > 3$).

I relate entanglement and separabilty properties of n-qubits systems to HS decompositions. Such decompositions relate the correlations of 2-qubits systems to certain one and two qubits measurements, the correlations for 3-qubits systems to certain one, two and three qubits measurements, for 4-qubits system to one, two, three, and four qubits measurements, etc. The multiplications in HS decompositions are, however, over Pauli matrices, which are not density matrices, and therefore are not related directly to entanglement. I show in the present work a special method by which density matrices for 2-qubits, 3-qubits and 4-qubits systems, related to HS decompositions, are quantified for entanglement and separability, by using equations (1), (2), etc. The advantage of the present method is that it can be generalized to large n-qubits systems ($n > 2$), and is valid for both pure and mixed states.

## 2. Separability and entanglement of 2-qubits systems related to HS decompositions

The mean value of Hermitian operator $O$, in an Hilbert Space $H$ described by a density matrix $\rho$, is given by

$$\langle O \rangle = Tr(\rho O) \quad . \tag{3}$$

We can find a set of operators $O$ so that (3) can be solved, uniquely for $\rho$. In order to determine the density operator of n-qubits system, $2^{2n}-1$ real numbers are required (because $\rho$ is Hermitian, satisfying $Tr\rho = 1$ ). We need therefore $2^{2n}-1$ independent observable by which we can determine $\rho$ (and for pure states $\rho^2 = \rho$ ).

For 2-qubits system denoted by A and B, we use the HS representation of the density matrix [11]:

$$4\rho = (I)_A \otimes (I)_B + (\vec{r}\cdot\vec{\sigma})_A \otimes (I)_B + (I)_A \otimes (\vec{s}\cdot\vec{\sigma})_B + \sum_{m,n=1}^{3} t_{nm} (\sigma_n)_A \otimes (\sigma_m)_B \quad . \tag{4}$$

Here $I$ denotes a unit matrix of dimension $2\times 2$ , $\otimes$ denotes outer product, the three Pauli matrices are represented by $\vec{\sigma}$ , while $\vec{r}$ and $\vec{s}$ represent 3-dimensional parameters-vectors, and summations over Pauli matrices are given by $m,n=1,2,3$ . We find that a pure 3-qubits entangled state is described by 15 parameters: 6 for $\vec{r}$ and $\vec{s}$ , and 9 for $t_{mn}$ .

These coefficients can be obtained by tracing the multiplication of $\rho$ by the corresponding term in (4). For example:

$$t_{nm} = Tr\{\rho(\sigma_n)_A \otimes (\sigma_m)_B\} \quad . \tag{5}$$

In order to obtain $r_i(A)$ or $s_j(B)$ (i, j =1,2,3) one needs to perform one measurement on a corresponding arm of a measuring device, while the parameters $t_{mn}$ , which might involve possible quantum correlations, are obtained by performing measurements on two arms.

By assuming the conditions $\vec{r} = \vec{s} = 0$ in (4), we get the density matrix:

$$4\rho_{AB} = \left\{ (I)_A \otimes (I)_B + \sum_{m,n=1}^{3} t_{nm} (\sigma_n)_A \otimes (\sigma_m)_B \right\} \tag{6}$$

I find that the 2-qubits Bell states and the special Werner state analyzed in [2] are described by density matrices which are of this form. In order to find the entanglement properties of the density matrix (6), I define a matrix $S_{AB}$ by:

$$
\begin{aligned}
4S_{AB} &= \sum_{m,n=1}^{3} t_{nm} \left(\sigma_n\right)_A \otimes \left(\sigma_m\right)_B + \left(I\right)_A \otimes \left(I\right)_B \sum_{m,n=1}^{3} \left|t_{n.m}\right| = \\
&\sum_{m,n=1}^{3} \frac{\left|t_{nm}\right|}{2} \left[ \left(I-\sigma_n\right)_A \otimes \left(I - sign\left(t_{nm}\right)\sigma_m\right)_B + \left(I+\sigma_n\right)_A \otimes \left(I + sign\left(t_{nm}\right)\sigma_m\right)_B \right]
\end{aligned} \quad . \tag{7}
$$

We defined here for positive $t_{nm}$ (negative $t_{nm}$) $sign\left(t_{nm}\right)=1$ $\left(sign\left(t_{nm}\right)=-1\right)$. Each multiplication in the square brackets on the right side of (7) is proportional to a separable density matrix. $\left|t_{nm}\right|$, is the probability for each separable density matrix. The sum of these probabilities in the general case is, however, not equal to 1. The present idea is to study the entanglement properties by comparing the matrix (6) (including multiplications of Pauli matrices which are not density matrices) with the matrix $S_{AB}$ of (7). The difference between the density matrices $\rho_{AB}$ and $S_{AB}$ is given by

$$
4\rho_{AB} - 4S = \left[ \left(I\right)_A \otimes \left(I\right)_B \right] \left( 1 - \sum_{m,n=1}^{3} \left|t_{n.m}\right| \right) \quad . \tag{8}
$$

We can complement (7) with (8), obtaining the matrix $\rho_{AB}$ by

$$
4\rho_{AB} = 4S + \left[ \left(I\right)_A \otimes \left(I\right)_B \right] \left( 1 - \sum_{m,n=1}^{3} \left|t_{n.m}\right| \right) \quad . \tag{9}
$$

We find that $\rho_{AB}$ of (9) is separable density matrix under the condition

$$
\sum_{m,n=1}^{3} \left|t_{n.m}\right| \leq 1 . \tag{10}
$$

Eq. (10) gives a sufficient condition for separability for the density matrix (6), but not a necessary condition. The condition for separability for the density matrix (6) can be improved by using singular value decomposition (SVD) which transforms the matrix $t_{n.m}$ into diagonal form: $t_{i.i} \equiv t_i$. By using an additional analysis it has been shown [12] that a sufficient and necessary condition for separability for the density matrix (6) is given by

$$\sum_{m,n=1}^{3} |t_i| \leq 1 \qquad . \qquad (11)$$

Assuming $t_{nm} = 0 \ (m,n = 1,2,3)$, $\vec{s} = 0$ in (4), we get the density matrix corresponding to measurement of qubit A in one arm of a measuring device:

$$\rho_A = \frac{1}{4}\left[(I)_A \otimes (I)_B + (\vec{r}\cdot\vec{\sigma})_A \otimes (I)_B\right] = \frac{1}{4}\left\{\left[(I)_A + (\vec{r}\cdot\vec{\sigma})_A\right] \otimes (I)_B\right\} \quad . \qquad (12)$$

Assuming $t_{nm} = 0 \ (m,n = 1,2,3)$, $\vec{r} = 0$ in (4), we get the density matrix corresponding to measurement of qubit B in the other one arm:

$$4\rho_B = \left[(I)_A \otimes (I)_B + (I)_A \otimes (\vec{s}\cdot\vec{\sigma})_B\right] = \left\{(I)_A \otimes \left[(I)_B + (\vec{s}\cdot\vec{\sigma})_B\right]\right\} \quad . \qquad (13)$$

We find according to (12-13) that $\rho_A$ and $\rho_B$ are separable density matrices not containing any quantum correlations.

The separability conditions for the density density matrix (4) have been studied by the use of Lorentz transformations [12-14].

## 3. Separability and entanglement properties of 3 qubits systems related to HS parameters

For 3-qubits system denoted by A, B and C we use the HS representation of this density matrix given by:

$$\rho = \frac{1}{8}\left\{\begin{array}{l} (I)_A \otimes (I)_B \otimes (I)_c + (\vec{r}\cdot\vec{\sigma})_A \otimes (I)_B \otimes (I)_c + (I)_A \otimes (\vec{s}\cdot\vec{\sigma})_B \otimes (I)_C + (I)_A \times (I)_B \otimes \left(\vec{f}\cdot\vec{\sigma}\right)_C \\ + \sum_{m,n=1}^{3} t_{mn} (I)_A \otimes (\sigma_m)_B \otimes (\sigma_n)_C + \sum_{k,l=1}^{3} o_{kl} (\sigma_k)_A \otimes (I)_B \otimes (\sigma_l)_C \\ + \sum_{i,j=1}^{3} p_{ij} (\sigma_i)_A \otimes (\sigma_j)_B \otimes (I)_C + \sum_{\alpha,\beta,\gamma=1}^{3} G_{\alpha\beta\gamma} (\sigma_\alpha)_A \otimes (\sigma_\beta)_B \otimes (\sigma_\gamma)_C \end{array}\right\}$$

$$. \qquad (14)$$

Here $I$ denotes the unit $2\times 2$ matrix, $\otimes$ denotes outer product, the three Pauli matrices are represented by $\vec{\sigma}$ , $\vec{r},\vec{s}$ and $\vec{f}$ are 3-dimensional parameters vectors, and summations over Pauli matrices are over $(m,n,k,l,i,j,\alpha,\beta,\gamma=1,2,3)$ . We find that a pure 3-qubits entangled state is described by 63 parameters: 9 for $\vec{r},\vec{s}$ , and $\vec{f}$ , 27 parameters for $t_{mn},o_{kl}$ , and $p_{ij}$ , and 27 $G_{\alpha\beta\gamma}$ parameters. These parameters can be obtained by tracing the multiplication of $\rho$ by the corresponding term in (14). For example:

$$G_{\alpha\beta\gamma}=Tr\left\{\rho\left(\sigma_{\alpha}\right)_{A}\otimes\left(\sigma_{\beta}\right)_{B}\otimes\left(\sigma_{\gamma}\right)_{C}\right\} \quad . \qquad (15)$$

I find that the parameters $\vec{r},\vec{s}$ and $\vec{f}$ are obtained by measurements in one arm of a measurement device, $t_{mn},o_{kl}$ and $p_{ij}$ , are obtained by measurement in the corresponding two arms of the measuring device, and $G_{\alpha\beta\gamma}$ are obtained by the corresponding measurements in the three arms of the measurement device.

Since we are interested in qubits systems which can give quantum correlations by measuring all qubits of the system, I will analyze, first, in the present Section a 3-qubits system with HS parameters corresponding only to 3-qubits measurement. For this special system we assume in (19): $\vec{r}=\vec{s}=\vec{f}=0$, and $t_{mn}=o_{kl}=p_{ij}=0$ for any two-qubits parameters. Then, we get the following density matrix:

$$8\rho_{ABC}=\left\{\left(I\right)_{A}\otimes\left(I\right)_{B}\otimes\left(I\right)_{c}+\sum_{\alpha,\beta,\gamma=1}^{3}G_{\alpha\beta\gamma}\left(\sigma_{\alpha}\right)_{A}\otimes\left(\sigma_{\beta}\right)_{B}\otimes\left(\sigma_{\gamma}\right)_{C}\right\} \quad . \qquad (16)$$

It can be shown that the partial transposition (PT) given by Peres [9] when it operates on the density matrix (16) does not change its eigenvalues so that it cannot give information on entanglement [15]. We can, however, can get information about separability using our methods. In order to find the separability properties of (16), a matrix $S_{ABC}$ is defined by:

$$8S_{ABC}=\sum_{\alpha,\beta,\gamma=1}^{3}G_{\alpha\beta\gamma}(\sigma_\alpha)_A\otimes(\sigma_\beta)_B\otimes(\sigma_\gamma)_C+(I)_A\otimes(I)_B\otimes(I)_C\left(\sum_{\alpha,\beta,\gamma=1}^{3}\left|G_{\alpha,\beta,\gamma}\right|\right)=(1/4)\sum_{\alpha,\beta,\gamma=1}^{3}\left|G_{\alpha,\beta,\gamma}\right|\cdot$$

$$\left\{\begin{array}{l}\left[\left\{(I)_A+(\sigma_\alpha)_A\right\}\otimes\left\{(I)_B-(\sigma_\beta)_B\right\}\otimes\left\{(I)_C-sign\left(G_{\alpha\beta\gamma}\right)(\sigma_\gamma)_C\right\}\right]+\\ \left[\left\{(I)_A+(\sigma_\alpha)_A\right\}\otimes\left\{(I)_B+(\sigma_\beta)_B\right\}\otimes\left\{(I)_C+sign\left(G_{\alpha\beta\gamma}\right)(\sigma_\gamma)_C\right\}\right]+\\ \left[\left\{(I)_A-(\sigma_\alpha)_A\right\}\otimes\left\{(I)_B-(\sigma_\beta)_B\right\}\otimes\left\{(I)_C+sign\left(G_{\alpha\beta\gamma}\right)(\sigma_\gamma)_C\right\}\right]+\\ \left[\left\{(I)_A-(\sigma_\alpha)_A\right\}\otimes\left\{(I)_B+(\sigma_\beta)_B\right\}\otimes\left\{(I)_C-sign\left(G_{\alpha\beta\gamma}\right)(\sigma_\gamma)_C\right\}\right]\end{array}\right\}$$

(17)

One should notice that the multiplications, in each squared brackets of (17), are proportional separable density matrix. $\left|G_{\alpha,\beta,\gamma}\right|$ , might be considered as a probability but in the general case $\sum_{\alpha,\beta,\gamma=1}^{3}\left|G_{\alpha,\beta,\gamma}\right|$ is not equal to 1. Here again, we study the separabilty properties of $\rho_{ABC}$, by comparing (16) with the matrix $S_{ABC}$ given by (17). The difference between the density matrices $\rho_{ABC}$ and and the matrix $S_{ABC}$ is given by

$$8\rho_{ABC}-8\tilde{\rho}_{ABC}=\left[(I)_A\otimes(I)_B\otimes(I)_C\right]\left(1-\sum_{\alpha,\beta,\gamma=1}^{3}\left|G_{\alpha,\beta,\gamma}\right|\right) \quad . \tag{18}$$

We find that $\rho_{ABC}$ of (18) is a separable density matrix under the condition

$$\sum_{\alpha,\beta,\gamma=1}^{3}\left|G_{\alpha,\beta,\gamma}\right|\leq 1 \quad . \tag{19}$$

Eq. (19) gives a sufficient condition for separabilty but it is not a necessary condition. This condition for separability can be improved by using analytical methods related to high order SVD [15]. The expression $\sum_{\alpha,\beta,\gamma=1}^{3}\left|G_{\alpha,\beta,\gamma}\right|$ can be referred as the $l_1$ norm, and this norm can be minimized under orthogonal transformations as this norm is not invariant under orthogonal transformations [15].

## 4. Entanglement properties of GHZ and $|B_i\rangle$ 3-qubits states related to HS decompositions

We denote in the standard basis $|0\rangle = \begin{pmatrix} 1 \\ 0 \end{pmatrix}$ and $|1\rangle = \begin{pmatrix} 0 \\ 1 \end{pmatrix}$ as the two states of each qubit. Then, the computational basis of n-qubits states is given by $|Ci\rangle \quad (i = 1, 2, \cdots, n)$ where $|Ci\rangle$ can be described by n-dimensional vectors where in the i'th entry we have 1 and all other entries we have zero. In the standard basis, the density matrix of (4) is described by $4 \times 4$ matrix. A GHZ entangled state [16-17] can be described as:

$$|\psi\rangle = \frac{|1\rangle_A |1\rangle_B |1\rangle_C + |0\rangle_A |0\rangle_B |0\rangle_C}{\sqrt{2}} \quad . \tag{20}$$

By using the density matrix for this state, and (14) we get by straight forward calculations [11]:

$$G_{122} = G_{212} = G_{221} = -1 \quad ; \quad t_{33} = o_{33} = p_{33} = G_{111} = 1 \quad . \tag{21}$$

All other parameters are equal to zero. I find that the quantum correlations in the GHZ state are included in the $G_{\alpha\beta\gamma}$ parameters which are obtained, by measuring all qubits of the system and in the parameters $t_{33}, o_{33}, p_{33}$ obtained by measurements in two arms of the device. GHZ states are entangled as by operating on their density matrices with PT transformations [9] we get negative eigenvalues. The density matrices of the GHZ states can be considered as maximally entangled density matrices as by tracing these density matrices over qubit(s) we get completely separable density matrices. We show here another set of maximally entangled states which, so far, has not been treated by other authors.

We can use for qubits systems the group $B_n$ discovered by Artin [18, 19], where the generator operators $g_1, g_2, \cdots, g_{n-1}$ satisfy the $B_n$ Braid-group relations:

$$B_n = \left\langle \begin{matrix} g_1, g_2, \cdots, g_{n-1} \mid g_i g_j = g_j g_i & |i-j| > 1 \; ; \\ g_i g_j g_i = g_j g_i g_j \quad |i-j| = 1 \quad ; & g_i g_i^{-1} = \mathrm{I} \end{matrix} \right\rangle \quad . \tag{22}$$

For the $B_n$ operators $g_1, g_2, \cdots, g_{n-1}$, operating in our system on the n-qubits system, we use the representation:

$$
\begin{aligned}
&g_1 = R \otimes I \otimes I \cdots \otimes I \quad ; \\
&g_2 = I \otimes R \otimes I \cdots \otimes I \quad ; \\
&g_3 = I \otimes I \otimes R \cdots \otimes I \quad ; \\
&\cdot \\
&\cdot \\
&\cdot \\
&g_{n-1} = I \otimes I \otimes I \cdots \otimes I \otimes R
\end{aligned} \quad . \tag{23}
$$

Here $\otimes$ denotes outer product, $R$ is the unitary matrix given for our system by [20]:

$$
R = \frac{1}{\sqrt{2}} \begin{pmatrix} 1 & 0 & 0 & 1 \\ 0 & 1 & -1 & 0 \\ 0 & 1 & 1 & 0 \\ -1 & 0 & 0 & 1 \end{pmatrix} \quad . \tag{24}
$$

The matrix $R$ satisfies also a special Yang-Baxter equation given by:

$$
(R \otimes I) \cdot (I \otimes R) \cdot (R \otimes I) = (I \otimes R) \cdot (R \otimes I) \cdot (I \otimes R) \quad . \tag{25}
$$

Here the dot represents ordinary matrix multiplication, and where $(R \otimes I)$ and $(I \otimes R)$ are matrices of $8 \times 8$ dimensions. The unitary $R$ matrix can be considered in quantum computation as a universal gate related to the CNOT gate by local single qubits transformations. It is quite easy to verify that the group relations (22) are satisfied by using (23-24).

The idea of using the $B_n$ operators [21] is that by operating with the multiplication operator $g_1 \cdot g_2 \cdot g_3 \cdots g_{n-1}$ on the computational n-qubits states $|Ci\rangle \quad (i = 1, 2, \cdots, n)$ we will get entangled states $|B_i\rangle \quad (i = 1, 2, \cdots, n)$:

$$
|B_i\rangle = g_1 \cdot g_2 \cdots g_{n-1} |Ci\rangle \quad . \tag{26}
$$

These states include entanglement properties analogous to those of the GHZ states [21]. One should take into account that $g_i \quad (i = 1, 2, \cdots, n-1)$ in (26), operate consequently on the qubits $n, n-1 \cdots, 2, 1$, and the matrix $R$ produces, entanglement between qubits pairs $(n-1, n) \cdots (2,3), (1,2)$ ,consequently.

For getting these general Bell entangled states we operate on computational states of the 3-qubits states with the operator

$$g_1 \cdot g_2 = (R \otimes I) \cdot (I \otimes R) \quad . \tag{27}$$

Similarly get the 4-qubits general entangled Bell states, by operating on computational states of the 4-qubits with the operator

$$g_1 \cdot g_2 . g_3 = (R \otimes I \otimes I) \cdot (I \otimes R \otimes I) \cdot (I \otimes I \otimes R) \quad . \tag{28}$$

In this way we get the present general $|B_i\rangle$ Bell entangled states for 3-qubits, 4-qubits, etc. systems. One should take into account that $g_1 \cdot g_2$ is a matrix of $8 \times 8$ dimension operating on 8 dimensional vectors, while $g_1 \cdot g_2 \cdot g_3$ is a matrix of $16 \times 16$ dimension operating on 16 dimensional vectors of the computational basis of states.

The explicit expression for the 3-qubits $|Bi\rangle \, (i = 1, 2, \cdots, 8)$ states have been derived in [21] and are given in (29). One can easily check that the quantum states $|B1\rangle, |B2\rangle, \cdots, |B8\rangle$ form another orthonormal basis of states for the 3-qubits system which have special entanglement properties. Any one of the 3-qubits $B_n$ entangled states includes a superposition of 4 multiplications with equal probability (i.e., real amplitudes are either 1 or -1), and each state in the same multiplication belongs to a different qubit. This property is analogous to the 2-qubit Bell states property, where we have two multiplications with equal probability and the first and second state in each Bell state belong to the first and second qubit, respectively.

$$
\begin{aligned}
2|B1\rangle &\equiv |0\rangle_A|0\rangle_B|0\rangle_C - |0\rangle_A|1\rangle_B|1\rangle_C - |1\rangle_A|0\rangle_B|1\rangle_C - |1\rangle_A|1\rangle_B|0\rangle_C \quad ; \\
2|B2\rangle &\equiv |0\rangle_A|0\rangle_B|1\rangle_C + |0\rangle_A|1\rangle_B|0\rangle_C + |1\rangle_A|0\rangle_B|0\rangle_C - |1\rangle_A|1\rangle_B|1\rangle_C \quad ; \\
2|B3\rangle &\equiv -|0\rangle_A|0\rangle_B|1\rangle_C + |0\rangle_A|1\rangle_B|0\rangle_C + |1\rangle_A|0\rangle_B|0\rangle_C + |1\rangle_A|1\rangle_B|1\rangle_C \quad ; \\
2|B4\rangle &\equiv |0\rangle_A|0\rangle_B|0\rangle_C + |0\rangle_A|1\rangle_B|1\rangle_C + |1\rangle_A|0\rangle_B|1\rangle_C - |1\rangle_A|1\rangle_B|0\rangle_C \quad ; \\
2|B5\rangle &\equiv -|0\rangle_A|0\rangle_B|1\rangle_C - |0\rangle_A|1\rangle_B|0\rangle_C + |1\rangle_A|0\rangle_B|0\rangle_C - |1\rangle_A|1\rangle_B|1\rangle_C \quad ; \\
2|B6\rangle &\equiv |0\rangle_A|0\rangle_B|0\rangle_C - |0\rangle_A|1\rangle_B|1\rangle_C + |1\rangle_A|0\rangle_B|1\rangle_C + |1\rangle_A|1\rangle_B|0\rangle_C \quad ; \\
2|B7\rangle &\equiv |0\rangle_A|0\rangle_B|0\rangle_C + |0\rangle_A|1\rangle_B|1\rangle_C - |1\rangle_A|0\rangle_B|1\rangle_C + |1\rangle_A|1\rangle_B|0\rangle_C \quad ; \\
2|B8\rangle &\equiv |0\rangle_A|0\rangle_B|1\rangle_C - |0\rangle_A|1\rangle_B|0\rangle_C + |1\rangle_A|0\rangle_B|0\rangle_C + |1\rangle_A|1\rangle_B|1\rangle_C \quad .
\end{aligned} \tag{29}
$$

Let us present the HS decompositions for some $|Bi\rangle$ entangled states by the use of (14):

$$
\begin{aligned}
&For \quad |B1\rangle: \quad G_{113} = G_{311} = G_{131} = -1 \quad ; \quad t_{22} = o_{22} = p_{22} = G_{333} = 1 \quad ; \\
&For \quad |B3\rangle: \quad G_{333} = G_{311} = G_{131} = t_{22} = o_{22} = -1 \quad ; \quad G_{113} = p_{22} = 1 \quad ; \\
&For \quad |B7\rangle: \quad t_{22} = G_{131} = p_{22} = -1 \quad ; \quad 0_{22} = G_{333} = G_{113} = G_{311} = 1 \quad .
\end{aligned} \tag{30}
$$

I find the interesting point that that the absolute values of these parameters are the same for all pure 3-qubits $|Bi\rangle$ entangled states, and only some of the signs are different. Out of the 63 HS parameters only 7 HS parameters are different from zero. The correlations for the $|Bi\rangle$ states of a 3-qubits system are similar to those of the GHZ states.

## 5. Conclusion

Separability, and entanglement for large n-qubits systems (n>2) can be quantified by using HS decompositions, following the present methods. For 2-qubits systems the present analysis is in agreement with previous analyses. We have given explicit calculations for 3-qubits systems including the GHZ and the present general Bell states referred as the $|Bi\rangle$ states. The present methods can be generalized in a straight forward way to larger n-qubits system (n>3).

## References

[1] M.A. Nielsen and I.L. Chuang. *Quantum Computation and Quantum Information*. Cambridge University Press, Cambridge, 2000.

[2] G. Benenti , G.Casati and G. Strini. *Principles of Quantum Computation and information. Volume I: Basic*

*concepts*. World -Scientific, Singapore, 2007.

[3] J. Audretsch. *Entangled systems*, Wiley, Berlin, 2007.

[4] W.K. Wootters. Entanglement of formation and concurrence. *Quantum Information and Computation* 1: 27-44, 2001.

[5] W.K. Wootters. Entanglement of information of an arbitrary state of two qubits. *Phys. Rev. Lett.* 80:2245-2248 (1998).

[6] C.H. Bennet , H.J. Bernstein, S. Popescu and B. Shumacher. Concentrating partial entanglement by local operations. *Phys. Rev. A* 53: 2046-2052, 1996.

[7] V. Coffman , J. Kundu and W.K. Wootters. Distributed entanglement. *Phys. Rev. A.* 61**:** 052306, 2000.

[8] I. Bengtsoon and K. Zyczkowski. *Geometry of quantum states*. Cambridge University Press, Cambridge,2008.

[9] A. Peres. Separability criterion for density matrices. *Phys. Rev. Lett.* 77: 1413-1415, 1996.

[10] M. Horodecki , P. Horodecki and R. Horodecki. Separability of mixed states: necessary and sufficient Conditions. *Phys. Lett. A* 223: 1-8, 1996.

[11] Y.Ben-Aryeh, A. Mann and B.C. Sanders. Empirical state determination of entangled two-level systems and its relation to information theory. *Foundations of Physics* 29: 1963-1975, 1999.

[12] Y.Ben-Aryeh and A.Mann, Explicit constructions of all separable two-qubits density matrices and related problems for three-qubits systems. *International Journal of Quantum Information* 13: 1550061 , 2015.

[13] F.Verstraete, K.Audenaert and B.De Moor. Maximally entangled mixed states of two qubits. *Phys. Rev. A* 64: 012316 (2001).

[14] F. Verstraete, J. Dehaene and B. De Moor. Local filtering operations on two qubits. *Phys. Rev. A* 64: 010101 (2001).

[15] Y.Ben-Aryeh and A. Mann, to be published.

[16] D.M. Greenberger, M.A. Horn and A. Zeilinger. Going beyond Bell's theorem. in *Bell's Theorem, Quantum Theory, and Conceptions of the universe pp. 69-72* edit. M. Kafatos M , Kluwer, Dordrecht, 1989.

[17] D. Bouwmeester , J-W. Pan, M.Daniell, H. Weinfurter and A. Zeilinger . Observation of three-photons Greenberger- Horne-Zeilinger entanglement. *Phys. Rev. Let.* 82: 1345-1349, 1999.

[18] J. Birman . Braids, links and mapping class groups. *Annals of Mathematics studies, No. 82* , Princeton University Press, Princeton, 1974.

[19] E. Artin. Theory of Braids. *Annals of Mathematics* 48 : 101-126, 1947.

[20] L.H. Kauffman and S.J. Jr. Lomonaco. Braiding operators are universal quantum gates. *New Journal of Physics.* 6,134:1-40, 2004.

[21] Y. Ben-Aryeh, Entangled states implemented by $B_n$ group -operators, including properties based on HS decompositions, separability and concurrence . *International Journal of Quantum Information* 13:1450045, 2015.